\documentclass[letterpaper,10pt]{article} 

\usepackage{opticameet3} 

\newcommand\authormark[1]{\textsuperscript{#1}}

\usepackage{amsmath,amssymb}
\usepackage[colorlinks=true,bookmarks=false,citecolor=blue,urlcolor=blue]{hyperref} 

\begin{document}

\title{Transformer-Based Prognostics: Enhancing Network Availability by Improved Monitoring of Optical Fiber Amplifiers}


\author{Dominic Schneider,\authormark{1,*} Lutz Rapp,\authormark{1} and Christoph Ament\authormark{2}}

\address{\authormark{1}Advanced Technology, Adtran Networks SE, 98617 Meiningen, Germany\\
\authormark{2}Faculty of Applied Computer Science, University of Augsburg, 86159 Augsburg, Germany}

\email{\authormark{*}dominic.schneider@adtran.com} 


\begin{abstract}
We enhance optical network availability and reliability through a lightweight transformer model that predicts optical fiber amplifier lifetime from condition-based monitoring data, enabling real-time, edge-level predictive maintenance and advancing deployable AI for autonomous network operation. \copyright~2025 The Author(s)
\end{abstract}

\section{Introduction}
Optical communication networks form the backbone of modern digital infrastructure, supporting high-speed connectivity and cloud-based services. Within these systems, optical fiber amplifiers (OFAs) are critical for maintaining signal integrity over long distances. Their degradation or failure can cause costly service disruptions and downtime, making predictive maintenance (PdM) essential for ensuring reliability and operational efficiency.

Traditional maintenance strategies, either reactive or scheduled, fail to meet the demands of large-scale optical systems. Reactive maintenance leads to unexpected service outages and costly emergency repairs, while scheduled interventions are blind to actual component condition and often result in unnecessary replacements or missed early failures. In contrast, data-driven prognostics, particularly remaining useful lifetime (RUL) forecasting, enable operators to anticipate component degradation, plan maintenance proactively, and reduce operational costs.

Predicting the RUL in OFAs is challenging due to nonlinear device behavior, feedback control dynamics, and limited monitoring data. To address these issues, we propose a lightweight model that combines structured sparse attention and low-rank parameterization to improve prediction accuracy while reducing computational complexity: the Sparse Low-Ranked Self-Attention Transformer (SLAT). Applied to OFA condition-based monitoring data, SLAT provides early, interpretable degradation insights and supports real-time, edge-level PdM deployment. Experiments on dedicated OFA datasets demonstrate that SLAT surpasses conventional deep learning models in accuracy and robustness, paving the way for scalable, AI-driven reliability management in optical networks.
\begin{figure}[bp]
  \label{approach}
  \centering
  \includegraphics[width=0.92\linewidth]{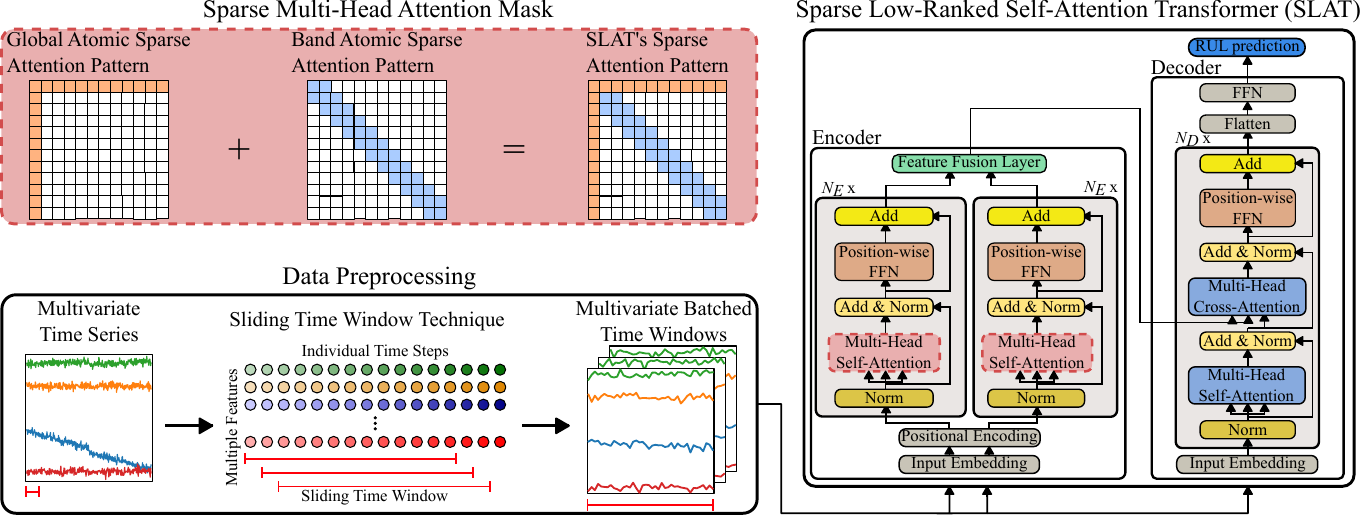}
  \caption{Proposed transformer architecture SLAT for predicting the RUL of OFAs, derived from multivariate time series data using sliding time window technique.}
\end{figure}

\section{Methodology and Experimental Setup}
The proposed framework combines structured data preprocessing with a tailored Transformer \cite{vaswani2017attention} architecture optimized for RUL forecasting of OFAs, as shown in Fig.~\ref{approach}. Condition-based monitoring (CBM) signals from the amplifier are segmented using a sliding time window approach, transforming raw sensor sequences $X=\{x_{1},x_{2},\dots,x_{T}\}$ into overlapping multivariate tensors of fixed length $N_{\text{stw}}$. Each window captures the short-term dynamics of the system and is enriched with statistical descriptors such as the mean and linear trend of each signal. These features are concatenated with normalized sensor data to improve degradation sensitivity while keeping input dimensionality consistent across datasets.

The Sparse Low-Ranked Self-Attention Transformer (SLAT) extends the standard encoder-decoder Transformer to efficiently handle multivariate time series under limited data conditions. Two parallel encoders independently process temporal and sensor-wise dependencies, and their outputs are fused before decoding. Each encoder block employs structured sparse attention, where each query token attends only to a restricted set of key tokens defined by a combination of global and local (banded) patterns:
\begin{equation}
  \text{Attention}\left(Q,K,V\right)=AV=\text{softmax}\left(\frac{M\odot\left(QK^\top\right)}{\sqrt{d_{k}}}\right)V,
\end{equation}
%
where $M$ is the binary sparse attention mask encoding global and band connections, and $\odot$ denotes the Hadamard product. This reduces quadratic complexity while preserving long- and short-range dependencies. To further regularize the model, the attention weight matrices $W_{Q},W_{K},W_{V}$ are low-rank parameterized as:
\begin{equation}
  Q=XW_{Q},\quad K=XW_{K},\quad V=XW_{V},
\end{equation}
%
with rank $r \ll d_{k}$, reducing parameter count and improving generalization in small-sample regimes.

Key hyperparameters of SLAT were optimized using Bayesian Optimization. The resulting configuration, summarized in Table~\ref{bayes}, includes four encoder blocks for both temporal and sensor paths, two decoder blocks, and eight attention heads per block, using 64-dimensional linear embeddings. This design achieves an effective trade-off between predictive precision and computational efficiency, making the model suitable for real-time deployment on distributed optical systems.

\begin{figure}[tbp]
  \label{setup}
  \centering
  \includegraphics[width=0.92\linewidth]{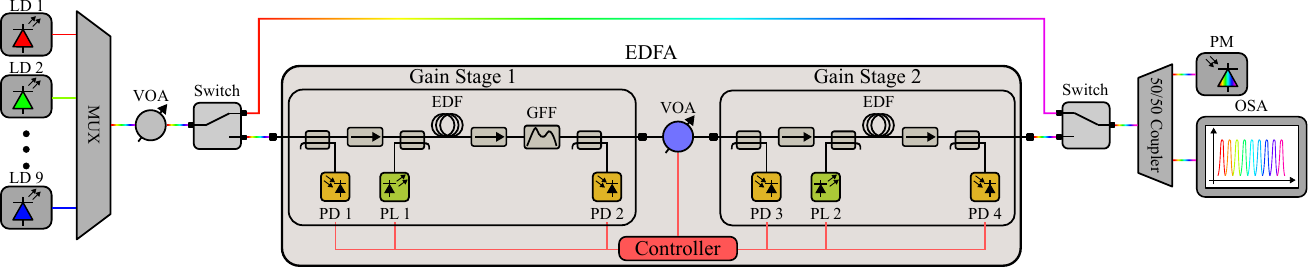}
  \caption{Data acquisition setup, illustrating the collection of multivariate time series data from OFAs, with induced degradation scenarios in the shown EDFA architecture.}
\end{figure}

Experimental data were obtained from a controlled OFA testbed, shown in Fig.~\ref{setup}, designed to emulate realistic degradation scenarios. The setup comprises a two-stage erbium-doped fiber amplifier (EDFA) with integrated sensors and actuators, including pump lasers (PL), power detectors (PD), a variable optical attenuator (VOA), and a gain-flattening filter (GFF).

\begin{table}[b]
    \begin{minipage}{.5\linewidth}
      \label{bayes}
      \caption{Bayesian-optimized Hyperparameters}
      \centering
      \begin{tabular}{lc}
        \hline\hline
        Hyperparameter & SLAT \\
        \hline
        Input embedding & 64 units, linear\\
        Sensor encoder & 4 blocks, 8 heads, GELU\\
        Time encoder & 4 blocks, 8 heads, GELU\\
        Decoder & 2 blocks, 8 heads, GELU\\
        Output layer & 1 unit, linear\\
        \hline\hline
      \end{tabular}
    \end{minipage}
    \begin{minipage}{.5\linewidth}
      \label{rmse}
      \caption{RMSE results for RUL prediction}
      \centering
      \begin{tabular}{lcccc}
        \hline\hline
        Subdataset & BiLSTM & DCNN & DAST & SLAT\\
        \hline
        PL & 10.18 & 9.72 & 9.66 & 8.93\\
        PD & 9.03 & 8.75 & 8.28 & 7.67\\
        VOA & 3.73 & 2.70 & 1.58 & 1.34\\
        PC & 10.81 & 10.20 & 9.99 & 8.29\\
        \hline
        Average & 7.72 & 7.85 & 7.50 & 6.56\\
        \hline\hline
      \end{tabular}
    \end{minipage}
\end{table}
A hardware-in-the-loop (HIL) simulator interfaces with the amplifier's embedded controller via SSH and Telnet to inject parameter drifts corresponding to physical degradation processes (e.g., pump aging or photodiode sensitivity loss). This allows precise, repeatable generation of CBM data across different fault modes and operating conditions. A detailed description of the test environment, instrumentation, and dataset structure is provided in \cite{schneider2025transformer}.

\section{Results}
To assess the forecasting performance of the proposed model SLAT, we compare its RUL forecasting accuracy against three established deep learning architectures: the Bidirectional Long Short-Term Memory (BiLSTM) network, the Deep Convolutional Neural Network (DCNN), and the Dual-Aspect Self-Attention Transformer (DAST) model \cite{kim2021multitask,wang2018remaining,zhang2022dual}. All models were trained and evaluated under identical experimental conditions.

Table~\ref{rmse} summarizes the root mean squared error (RMSE) obtained across all subdatasets representing soft-failures of the following components: pump laser (PL), power detector (PD), variable optical attenuator (VOA), and passive components (PC). Across all categories, SLAT consistently outperforms the state-of-the-art baselines, achieving the lowest average RMSE of 6.56, compared to 7.50 for DAST, 7.72 for BiLSTM, and 7.85 for DCNN. The results demonstrate that SLAT achieves a robust trade-off between predictive accuracy and computational efficiency, validating its suitability for real-time, distributed PdM in optical transport systems.

\begin{figure}[tbp]
  \label{rtf}
  \centering
  \includegraphics[width=\linewidth]{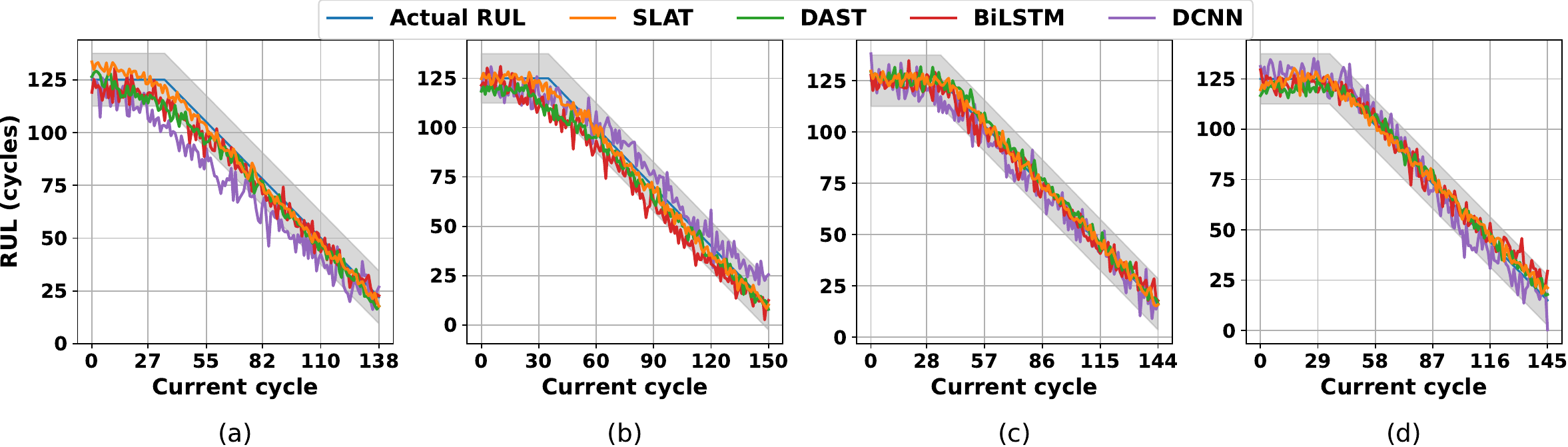}
\caption{RTF trajectories for representative test samples of each degradation scenario: (a) pump laser, (b) power detector, (c) variable optical attenuator, and (d) passive components.}
\end{figure}

To further examine model behavior over time, Run-to-Failure (RTF) trajectories were generated for representative test samples of each degradation scenario, as shown in Fig.~\ref{rtf}(a-d). Each subplot compares the predicted RUL from SLAT with the ground-truth trajectory and with the outputs of DAST, BiLSTM, and DCNN.

Across all scenarios, SLAT's predicted trajectories remain within a narrow confidence interval around the measured RUL, confirming both the accuracy and temporal stability of its forecasts. These visual results reinforce the quantitative findings summarized in Table~\ref{rmse} and illustrate SLAT's capability to model diverse optical degradation processes with high reliability.

\section{Conclusion}
We proposed a lightweight transformer model that enables accurate, real-time prediction of amplifier lifetime from monitoring data, enhancing overall optical network resilience and operational efficiency. With its low computational complexity, SLAT advances edge-intelligent, AI-driven maintenance toward fully autonomous optical networks.

\vspace{10pt}
\textbf{Acknowledgment} This work has received funding from the German Federal Ministry of Research, Technology, and Space (BMFTR) project SUSTAINET-Advance, Grant 16KIS2271K, in the framework of the CELTIC-NEXT project id C2024/3-3.

\bibliographystyle{opticajnl}
\bibliography{bibliography}

\end{document}